\def\lae{\mathrel{<\kern-1.0em\lower0.9ex\hbox{$\sim$}}}
\def\gae{\mathrel{>\kern-1.0em\lower0.9ex\hbox{$\sim$}}}

\newcommand{\be}{\begin{equation}}
\newcommand{\ee}{\end{equation}}

\documentclass[apj]{emulateapj}
\shorttitle{} \shortauthors{Fang \& Zhang}
\usepackage{apjfonts,lscape}

\begin{document}

\title{A Revisit of the Two-Pole Caustic Model for GeV Light Curves of Gamma-Ray Pulsars}
\author{J. Fang \& L. Zhang}
\affil{Department of Physics, Yunnan University, Kunming, China}
\email{lizhang@ynu.edu.cn}

\begin{abstract}

The GeV light curve of a pulsar is an important probe to detect
acceleration regions in its magnetosphere. Motivated by the recent
reports on the observations of pulsars by {\it Fermi} Large Area
Telescope (LAT), we restudy the two-pole caustic model and revise it
to investigate the properties of the light curves in the GeV band.
In the revised model, although acceleration gaps can extend from the
star surface to the light cylinder along near the last open field
lines, the extension of the gaps along the azimuthal direction is
limited because of photon-photon pair production process. In such
gaps, high-energy photons are emitted uniformly and tangentially to
the field lines but cannot be efficiently produced along these field
lines where the distances to the null charge surface are larger than
$\sim0.9$ times of the distance of the light cylinder, and the
effective azimuth extension of the gaps is about $230^\circ$. The
model is applied to the four pulsars Vela, PSR J1028-5819, PSR
J0205+6449, and PSR J2021+3651 whose light curves obtained with {\it
Fermi} have been recently released. The model is successful in
reproducing the general feature of the light curves for the four
pulsars, and the radial distances of the radio pulse for the four
pulsars are estimated.
\end{abstract}


%\newpage
\section{Introduction}
\label{sec:intro}

High-energy observations with the {\it Fermi} Large Area Telescope
in the GeV band of four pulsars, Vela, PSR J1028-5819, PSR
J0205+6449, and PSR J2021+3651 have been recently reported, and the
GeV light curves of the four pulsars show two main peaks separated
in phase by 0.4--0.5 and lagged the radio peak by a phase shift of
0.08--0.2\citep{A09a, A09b, A09c, A09d}. These new observed results
provide important limitations to $\gamma$-ray emitting models of
pulsars.

Many models have been proposed to explain high-energy properties of
$\gamma$-ray pulsars. The common feature of these models is that
electrons and/or positrons are accelerated by the electric field
parallel to magnetic field and produce photon emission through
different radiation processes in the open volume of the pulsar's
magnetosphere. On the other hand, the main difference of the these
models is that acceleration regions are different. For examples,
photon emission occurs within several radii of a neutron star above
a polar cap surface in the polar cap models
\citep[e.g.,][]{dh82,dh94,set95}. However, in the outer gap models
\citep[e.g.,][]{ry95,zc97,zc98,zet04,tea08}, both particle
acceleration and photon emission take place in the outer parts of
the pulsar's magnetosphere.

%a and two-pole caustic ones \citep[][]{dr03,drh04}.

In this paper, we focus on the interpretation of GeV light curves of
the $\gamma$-ray pulsars which involves three dimensional models for
$\gamma$-ray pulsars. In the polar cap models, the characteristic of
the two peaks in the light curve are usually produced when the line
of sight intersects the polar cap beam, and  a nearly aligned
rotator is needed to reproduce the observed properties of the light
curve \citep[e.g.,][]{dh94}. Based on the geometrical study of polar
cap accelerators, slot gap model in which particles continue to
accelerate and radiate to high altitude along the last open field
lines has been proposed \citep[e.g.,][]{Arons83,AS79,MH03} and
reasonably reproduces the Crab pulsar phase-resolved spectra
\citep{het08}. In the outer gap and the two-pole caustic models,
$\gamma$-rays are emitted in the acceleration gap located near the
last open field lines, and prominent peaks in the light curve are
produced by the caustic effects, i.e., the aberration of photon
direction and the time-of-flight delay caused by the finite speed of
light \citep{ry95,cet00,dr03,Dea04}. For a polar cap of a
$\gamma$-ray pulsar, high-energy photons emitted along the leading
field lines extend in phase, but those along the trailing lines can
be piled up in phase as a integrative result of the special
configuration of the trailing lines and the caustic effects.

In the original outer gap models, the gap starts from the null
charge surface to the light cylinder, with the last open field lines
as the lower boundary and a charge layer on a surface of the open
field lines as the upper boundary
\citep[e.g.,][]{cet86a,cet86b,cet00,ry95}. High energy photons
predicted by these models are produced beyond the null charge
surface, and then only these from one pole can be observed. However,
more recent electro-dynamical studies show that the gap position
shifts if particles can be injected either at the inner boundary or
at the outer one, and the gap can be located at the star surface if
the injection rate across the boundary is comparable to the
Goldreich-Julian value \citep[][]{hs01a,hs01b,hs02,hhs03,tsh04}.
\citet{tea08} proposed a revised outer-gap model to investigate the
multi-band phase-resolved spectra for the Crab pulsar. In the model,
the gap extends significantly towards the star since a current
carried by the pairs could be produced in the gap although ignoring
the current injection from the inner and the outer boundaries. As a
result, the emission from the both poles can contribute to the
observed light curves.

\citet{dr03} firstly proposed the two-pole caustic model to
interpret the high-energy light curves of pulsars. In their model,
photons are emitted uniformly in the gap, and the gap extending from
the star surface to high altitudes is confined to the last open
field lines. Two sharp peaks with well-developed wings and a phase
separation of 0.4--0.5 can be easily reproduced using the model;
moreover, a bridge emission and a significant off-pulse emission
also appear in the light curve. Therefore, the light curves of the
pulsars with two main peaks separated by 0.4--0.5 in phase can be
well explained using the two-pole caustic model.

Motivated by the recent reports on the observations with the {\it
Fermi} Large Area telescope for the four normal pulsars, we restudy
the high-energy light curves for these pulsars basically following
the two-pole caustic model \citep[][]{dr03,Dea04} and find that the
off-pulse emission level in the observed GeV light curves compared
with the resulting one with the two pole caustic model is very low,
especially, for Vela and PSR J0205+0449. On the other hand, after
taking the pair production process into account, \citet{cet00} found
that the distance to the null charge surface is the function of
azimuthal angle $\phi$ \citep[also see][]{zc01,tea08}, i.e. $r_{\rm
nul}=r_{\rm nul}(\phi)$ and $r_{\rm nul}(\phi>0^{\circ})>r_{\rm
nul}(\phi=0^{\circ})$, where $\phi=0^{\circ}$ represents the
magnetic meridional plane. In other words, the extension of the
outer gap along the azimuthal direction is limited by the pair
production process. Therefore, we revise the two-pole caustic model
as follows: high-energy photons cannot be efficiently produced along
the field lines where the distances to the null charge surface are
larger than $f_{\rm in}\sim1$ times of the distance of the light
cylinder, and we choose $f_{\rm in}=0.9$ in this paper. As a result,
the level of the emission of the light curve in the off-pulse region
can be greatly reduced, and the result is more consistent with the
observations.  We apply the model to the four pulsars, Vela, PSR
J1028-5819, PSR J0205+6449, and PSR J2021+3651 whose light curves
obtained with {\it Fermi} have been recently reported, and the
results of the model are consistent with the observations.

\section{Model and Results}
\label{sec:model}

A retarded vacuum dipole geometry of the magnetic field around the
pulsar is assumed in the model \citep[see details in][]{cet00}.
Although the full three-dimensional MHD solutions of the pulsar
magnetosphere can describe the structure of the magnetosphere filled
with charges more accurately \citep[][]{s06,kc09}, the simulations
are numerical and time-consuming. The retarded vacuum dipole which
can approximate the MHD solutions is analytical and easy to carry
out, thus we use the retarded vacuum dipole in the calculation
\citep[see also in][]{het08}. In the original two-pole caustic
model, the gap confined to the last open field lines is thin and
extends from the polar cap to the light cylinder; the high-energy
photons are emitted along the field lines extending from the star
surface to high altitudes and the emissivity is uniform within the
gap region \citep[][]{dr03,Dea04}.

The calculation basically follows the method in \citet{Dea04}.
Runge-Kugga integrations are employed to receive the shape of the
polar cap rim, and the open volume coordinates ($ r_{\rm ovc}$, $
l_{\rm ovc}$) are established to easily emulate the particle
distributions at the pulsar surface, where $ r_{\rm ovc} = 1 \pm
d_{\rm ovc}$, $d_{\rm ovc}$ is the minimum distance of a point from
the polar cap region in units of the standard polar cap radius, $
l_{\rm ovc}$ is the arc length along the deformed ring of the fixed
$r_{\rm ovc}$; the electron distribution is rim-dominated at the
star surface and can be expressed by a Gaussian function:
\begin{equation}
\frac{dN_{\rm ph}}{ds}\propto\exp \left(-\frac{(r_{\rm ovc} - r_{\rm
ovc}^0)^2}{2\sigma^2}\right), \label{equ:pd}
\end{equation}
here $\sigma$ describes the thickness of the gap; the emitting
region has an upper boundary $r_{\rm max}$, which is the distance of
the emitting region to the star, and the emissivity drops to zero at
a distance $\rho_{\rm max}=(0.75 - 0.95 R_{\rm lc})$ from the
rotational axis, where $R_{\rm lc}$ is the distance of the light
cylinder \citep[see details in][]{Dea04}. $r_{\rm ovc}^0=1$ means
that the particles emitting $\gamma$-rays are accelerated mainly
along the last open field lines. However, we use $r_{\rm
ovc}^0=0.98$ in the calculation as the physical outer-gap model in
which the surface containing the last open field lines is usually
treated as a boundary of the gap \citep[e.g.,][]{tea08}.

The high-energy photons are emitted tangentially to local magnetic
field lines in the corotating frame, and the direction $\eta$ in the
observer frame can be obtained from the direction $\eta '$ in the
corotating frame with
\begin{equation}
\eta = \frac{\eta' + [\gamma + (\gamma - 1)(\mathbf{\beta \cdot
\eta'})/\beta^2]\mathbf{\beta}}{\gamma (1 + \mathbf{\beta \cdot
\eta'})}, \label{equ:ste}
\end{equation}
where $\gamma=(1-\beta^2)^{-1/2}$, $\beta = \mathbf{v}/c$, ${\bf v}=
\mathbf{\Omega\times r}$, $\mathbf{\Omega}$ is the angular velocity
of the pulsar and $\mathbf{r}$ is the radial distance of the
emitting point \citep{dr03}. Then a phase of detection $\Phi$ can be
obtained with
\begin{equation}
\Phi = - \frac{\phi_{em}}{2\pi} - \frac{\mathbf{r\cdot \eta}}{2\pi
R_{\rm lc}}, \label{equ:phi}
\end{equation}
where $\phi_{em}$ is the azimuthal angle of $\eta$.

\begin{figure*}[]
\includegraphics[scale=1.35]{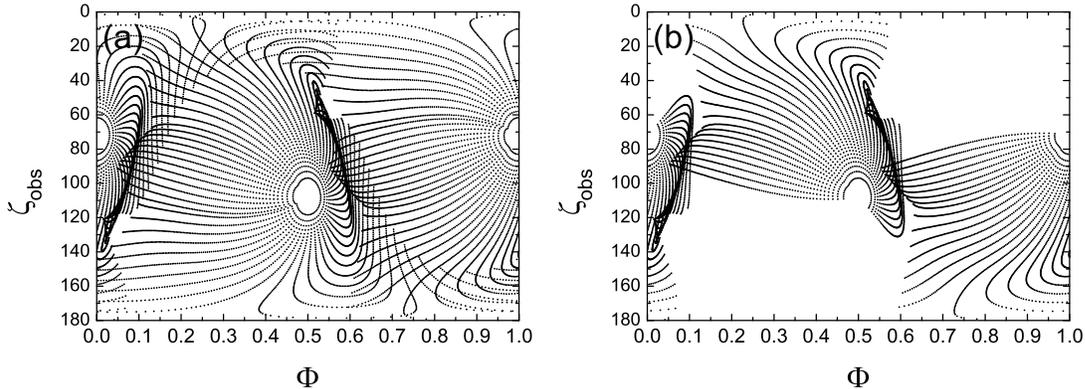}
\caption{\label{Figpmap} (a) Projection of the last open field lines
on the space ($\Phi, \zeta_{\rm obs}$). (b) Projection of the last
open field lines for which $r_{\rm nul}<0.9R_{\rm lc}$. The
parameters are $\alpha=72^{\circ}$, $r_{\rm max}=1.0R_{\rm lc}$,
$\rho_{\rm max}=0.95R_{\rm lc}$, and $P=89.3$ ms. }
\end{figure*}

The resulting projection of the last open field lines on the space
($\Phi, \zeta_{\rm obs}$) can be seen in panel (a) of
Fig.\ref{Figpmap}, where $\zeta_{\rm obs}$ is the viewing angle. The
two  blank deformed ovals correspond to the two polar caps, and two
caustics resulting from aberration and the finite speed of light can
be seen as the dark arches along the trailing field lines in the
projection map. So photons can be piled up along a broad range of
altitude because, for the trailing lines, the effect of different
altitudes on the final phase can be almost compensated by the
aberration and time of flight \citep[][]{m83,dr03,Dea04}, and two
main peaks can be usually seen in the pulsar high-energy light
curves.

\begin{deluxetable}{lccccc}
\tablecaption{\label{tab1} Parameters for the four pulsars}
\tablehead{
  \colhead{Name} & \colhead{P (ms)} & \colhead{$\alpha$} & \colhead{$\zeta$}
    & $\sigma$ & \colhead{$\delta_{\rm ph}$\tablenotemark{a}} }
\startdata
Vela         & 89.3  & 72  & 57  & 0.02 & 0.0425   \\
J2021+3651   & 103.7 & 70  & 81  & 0.02 & 0.065   \\
J1028-5819   & 91.4  & 75  & 65  & 0.02 & 0.12    \\
J0205+6449   & 65.7  & 85  & 78  & 0.04 & 0.01
\enddata
\tablenotetext{a}{A phase interval $\delta_{\rm ph}$ is added to the
resulting light curve to compare with the observation for each
pulsar.}
\end{deluxetable}

We now apply the model to the four normal $\gamma$-ray pulsars,
Vela, PSR J1028-5819, PSR J0205+6449, and PSR J2021+3651 whose GeV
light curves obtained with the {\it Fermi} have been recently
reported and have two main peaks separated by 0.4-0.5 in phase
\citep[][]{A09a,A09b,A09c,A09d}. Vela is a young pulsar with a
characteristic age of 11 kyr and a period of $\sim 89$ ms
\citep[][]{lea68}. The distance is $\sim 287$ pc from the VLBI
parallax measurement \citep[][]{dea03}. It is the brightest
persistent $\gamma$-ray source and was observed first during the
SAS2 mission \citep[][]{tea75} and followed by {\it COS B}
\citep[][]{kea80}, EGRET \citep[][]{kea94}, {\it AGILE}
\citep[][]{pea09} and {\it Fermi} \citep[][]{A09b}. The {\it Fermi}
telescope obtained 32,400 pulsed photons with energies $\geq0.03$
GeV , and the high-quality result shows that two main peaks, P1 and
P2, separated by $\sim 0.432$ in phase appear in the light curves
and a significant component in the bridge region \citep[][]{A09b}.
P1 lags the radio pulse at 1.4 GHz by a phase of $\sim 0.13$; the
two peaks are asymmetric, i.e., P2 has a slow rise and a fast fall,
whereas the fall of P1 is slower. Moreover, a third peak in the
bridge component becomes distinct at $>1$ GeV and the peak moves to
later phase with increasing energy\citep[][]{A09b}.

\begin{figure}
\includegraphics[scale=0.8]{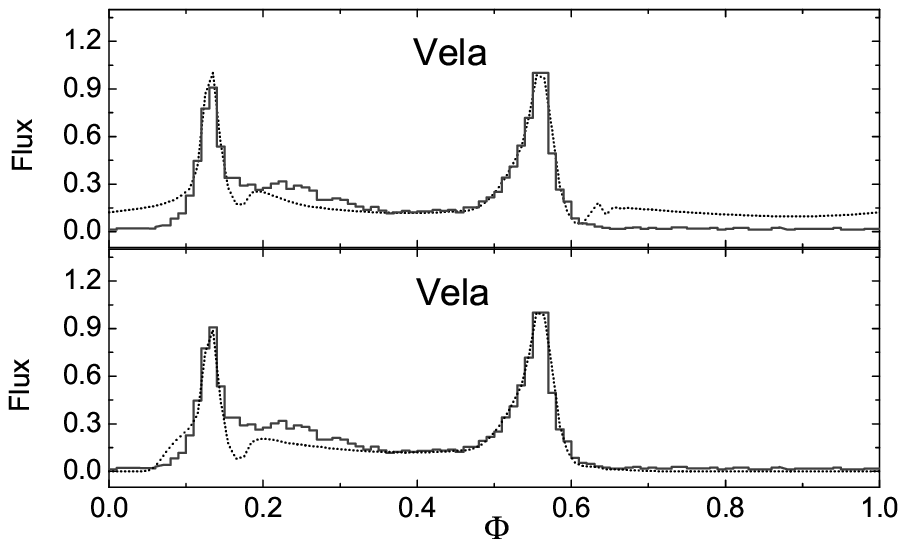}
\caption{\label{Figvela} The light curve for Vela (dotted line)
predicted by the model without (upper panel) and with (lower panel)
the prescription that the filed lines with $r_{\rm nul}>0.9R_{\rm
lc}$ cannot contribute emission to the resulting light curve. The
{\it Fermi} result (0.3 -- 1.0 GeV) \citep{A09b} is shown as solid
line for comparison. The parameters are shown in Table\ref{tab1}.}
\end{figure}

The resulting light curve for the pulsar with the model in this
paper can be seen in Fig.\ref{Figvela}, and the parameters are given
in Table\ref{tab1}. For Vela and the  other pulsars, the viewing
angle $\zeta_{\rm obs}$ is chosen basically according to the
available results deduced from the geometry of the pulsar wind
nebula in the X-ray band, and then we choose the other parameters to
well reproduce the {\it Fermi} light curves. Note that the phase 0
for the LAT data corresponds to the radio peak, whereas in this
paper it corresponds to photon emitted at the stellar center. If the
radio pulse is emitted above the stellar surface, the phase of the
radio peak will deviate from 0 due to the aberration and retardation
effects. in the resulting curve would coincident with the observed
ones assuming the radio pulse is the core component. Moreover, the
location of the radio emission for a pulsar can be 10 -- 1000 km
(\citealt{kj07,wea09}) high above the stellar surface. Therefore,
there is usually a phase difference between the resulting curve of
the model and the observed one with the LAT for the aberration and
retardation effects, and a phase interval $\delta_{\rm ph}$, which
is chosen according to the first peak in the resulting curve is
aligned with the observed one can be obtained for each pulsar.

Vela has a viewing angle of $\sim 63^\circ$ according to the torus
fitting procedure to the {\it Chandra} observation
\citep[][]{nr04,nr08}. We use different $\zeta_{\rm obs}$ around
this value and other appropriate parameters to detect which one can
well reproduce the observed profile. The final results show that the
profile of P1, P2 and the bridge emission in phase $\sim0.32 - 0.5$
can be well reproduced with the model and a notch following P1 also
appears in the resulting light curve. However, the model result
underestimates the emission in the P3 region ($\Phi=0.15-0.3$) and
significantly overestimates the emission in the off-pulse region
($\Phi = 0-0.1, 0.6-1.0$). As argued in \citet[][]{A09b}, the
emission in the P3 region is produced via the synchrotron emission
resulting from the relatively low-altitude pair cascades initiated
by the GeV curvature photons. The model in this paper assumes
photons to be emitted along the magnetic field lines with uniform
emissivity, so the model should not be used to tackle the P3 profile
although the notch following P1 can also be interpreted with it.

\begin{figure}
\includegraphics[scale=0.8]{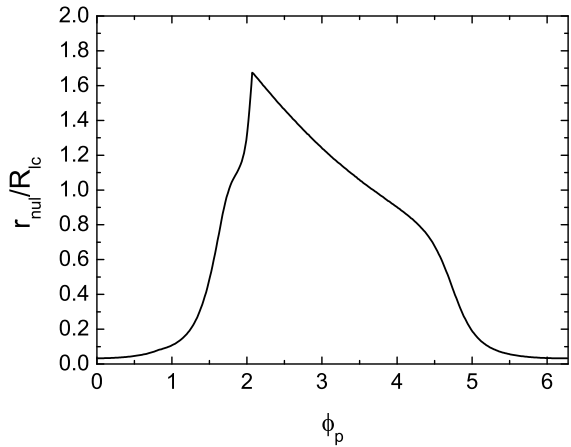}
\caption{\label{Rdis} Radial distance to the null charge surface for
the last open field lines, here $\phi_{\rm p}$ is the azimuth angle
the polar cap. The parameters are the same as Fig.\ref{Figpmap}.}
\end{figure}

Physically, in outer gap models when relativistic particles
propagate outwardly along the field lines in the gap of a pulsar,
some other particles with the opposite charge flowing towards the
star can emit high-energy $\gamma$-rays inwardly. In these models,
high-energy $\gamma$-rays can be emitted by the the
inwardly-propagating particles in the gap, and these photons can
collide with the X-rays from the star surface although the inward
emission is much fainter than the outward one. The inward emission
can usually be neglected when modeling the high-energy light curve
of a pulsar. The inner boundary of the gap in the outer gap models
is limited at the location where significant pairs are produced to
screen the gap \citep[][]{zc97,zc98,cet00,tea08}. Formerly, the
inner boundary of the gap is set to the null charge surface
\citep[e.g.,][]{cet00}. However, more recent detailed
electrodynamical studies show that the inner boundary can be shifted
from the null surface to the star surface with the current through
the gap taken into account \citep[][]{hs01a,tsh04,tea08}.
\citet[][]{tea08} used a prescription that $j_g(\phi)\infty r_{\rm
nul}(\phi)^{-3/8}$ neglecting the out current injected from the
inner and outer boundaries to get the location of the inner
boundary, where $j_g(\phi)$ is the current density in units of
$\Omega B/2\pi$ carried by the pairs created in the gap and $r_{\rm
nul}(\phi)$ is the radial distance to the null charge surface on the
last open field line for the polar cap azimuth angle $\phi$. The
pair creation process is sensitive to the gap geometry which is
still not accurately known \citep[][]{tea08}, and the
characteristics of the current through the gap in 3-dimensional
geometry are unclear now. Following \citet{tea08}, the altitude of
the inner boundary of the gap for each field line is proportional to
location of the null surface. So for the field lines with high null
surface, the gap cannot be formed or effective enough to produce
high-energy $\gamma$-rays. In fact, we found that if the emission
from the field lines with $r_{\rm nul}>0.9R_{\rm lc}$ is excluded,
the resulting light curve for Vela shows two peaks with zero flux
from the pulsar magnetosphere in the off-pulse region, which is more
consistent with the {\it Fermi} observation. The radial distance of
the null surface for each last open field line is shown in
Fig.\ref{Rdis}, and the azimuth extension of the last open field
lines emitting high-energy photons is $\sim 230^\circ$ for the Vela
parameters with $f_{\rm in}=0.9$.. Therefore, different from the
two-pole caustic model proposed by \citet{dr03} and \citet{dea04} in
which the particles can emit high-energy photons along all the last
open field lines, the high-energy photons from the field lines for
which $r_{\rm nul}$ is larger than $f_{\rm in}$ times of the
distance of the light cylinder are excluded. As a result, the
corresponding emission pattern of the lines is indicated in the
penal (b) of Fig.\ref{Figpmap}.

With the prescription that high-energy $\gamma$-rays cannot be
emitted along the lines for which $r_{\rm nul}>0.9R_{\rm lc}$, a
large part of area in the projection map is blank. In such a case,
the flux in the light curve following the P2 is greatly reduced with
$\alpha<80^\circ$ and just one main peak remains since the P1
disappears with $\alpha<40^\circ$; however, the profile of the
resulting light curve does not change for $\alpha\sim90^\circ$
because the filed lines with $r_{\rm nul}>0.9R_{\rm lc}$ do not
contribute emission to the light curve for this inclination angle.
The final light curve with the prescription taken into account is
shown in the lower panel of Fig.\ref{Figvela}. The flux outside the
two peaks is greatly reduced and the modeling light curve is more
consistent with the observation with a phase interval of 0.0425
added to the resulting light curve to make the phase of the first
peak consistent with the observed one.

\begin{figure}
\includegraphics[scale=0.8]{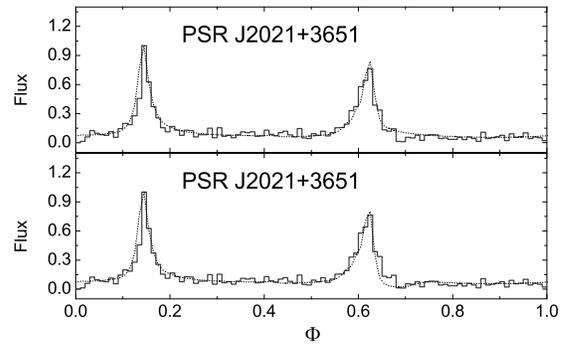}
\caption{\label{Fig2021} The light curve for PSR J2021+3651 (dotted
line) predicted by the model without (upper panel) and with (lower
panel) the prescription that the filed lines with $r_{\rm
in}>0.9R_{\rm lc}$ cannot contribute emission to the resulting light
curve. The {\it Fermi} result ($>0.1$ GeV) \citep{A09c} is shown as
solid line for comparison. Others are the same as
Fig.\ref{Figvela}.}
\end{figure}

PSR J2021+3651 is a young and energetic pulsar with a rotation
period of $\sim104$ ms and was discovered by \citet[][]{rea02} in
the radio search towards the five unidentified {\it ASCA} X-ray
sources coincident with the EGRET $\gamma$-ray sources. The pulsar
is associated with 3EG J2021-3716 \citep[][]{hea99,rea02}, and the
pulsed $\gamma$-rays were first discovered by \citet{hea08} with
{\it AGILE}. Possible pulsed X-rays from the pulsar and a pulsar
wind nebula (PWN) associated it had been detected using {\it
Chandra} \citep[][]{hea04}. The PWN has a "dragonfly" shape, and a
viewing angle of $86\pm1^\circ$ can be estimated for the pulsar from
a fit to the torus structure \citep[][]{vea08}. The high-resolution
$\gamma$-ray light curve of PSR J2021+3651 obtained with the Large
Area Telescope (LAT) on {\it Fermi} has been reported by
\citet{A09c}. Two narrow peaks separated by $\sim0.468$ in phase
appear in the curve, and the first peak lags the 2 GHz pulse by
$\sim0.162$ in phase \citep[][]{A09c}.

We model the GeV light curve of PSR J2021+3651 with the parameters,
$P=103.7$ ms $\alpha=70^\circ$, $\zeta_{\rm obs}=81^\circ$,
$\sigma=0.02$ (see Table\ref{tab1} and Fig.\ref{Fig2021}). For these
parameters, the flux in the off-pulse interval has a significant
amount, we choose 50 counts per 0.01 phase in the original observed
light curve \citep[][]{A09c} as the background emission level, i.e.,
the flux equaling 0 in Fig.\ref{Fig2021} corresponds to the
background emission 50 counts per 0.01 phase. A phase interval of
0.065 is chosen in the figure. The GeV light curve can be well
reproduced using the model, and, for $\alpha=70^\circ$, $\zeta_{\rm
obs}=81^\circ$, the light curve excluding the emission from the
field lines with $r_{\rm nul}>0.9R_{\rm lc}$ is nearly same as that
including the emission except a small notch around the phase 0.68.
Note that the a notch also appears in the observed light curve with
the LAT, which also sustains the correctness of our prescription.

\begin{figure}
\includegraphics[scale=0.8]{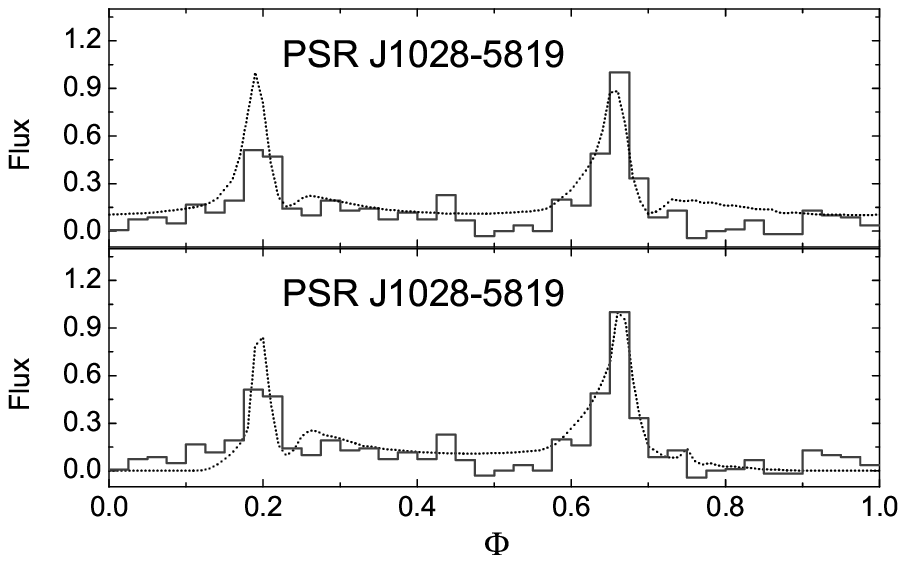}
\caption{\label{Fig1028} Comparison of the resulting light curve
with the {\it Fermi} observation ($>0.1$ GeV) \citep[][]{A09a} for
the pulsar PSR J1028-5819. Others are the same as
Fig.\ref{Figvela}.}
\end{figure}

The pulsar PSR J1028-5819, newly discovered in the high-frequency
search for pulsars using the Parkes telescope and the Australia
Telescope Compact Array on the unidentified EGRET source 3EG
J1027-5817, is a young pulsar with a period of 91.4 ms and a
characteristic age of $9.21\times10^4$ yr \citep[][]{kea08}. Pulsed
$\gamma$-ray signals from this pulsar have been recently discovered
by the LAT, and two narrow peaks P1 and P2 with a phase separation
of $\sim0.46$ appear in the light curve, i.e., P1 at phase
$\sim0.200$ and P2 at phase $\sim0.661$ with the phase 0 set to the
1.4 GHz radio pulse \citep[][]{A09a}.

The resulting light curves predicted by the model are shown in
Fig.\ref{Fig1028} with $P=91.4$ ms, $\alpha=75^\circ$, $\zeta_{\rm
obs}=65^\circ$ and $\sigma=0.02$ with a phase interval $\delta_{\rm
ph} = 0.12 $ is added to the resulting light curve to make the two
peaks consistent with the observation, which means the altitude of
the radio emission is $\sim263$ km. For these parameters, if all the
open field lines originating from the stellar surface at $r_{\rm
ovc0}$ can contribute to the observed light curve, the flux at P1 is
high above that at P2, and a significant amount of emission belongs
to the off-pulse interval; however, similar to the Vela pulsar, if
excluding the emission from the field lines with $r_{\rm
in}>0.9R_{\rm lc}$, the flux in the off-pulse region is greatly
reduced, and the flux at P1 is smaller than that at P2, which is
more consistent with the observation. Clearly from the lower panel
of Fig.\ref{Fig1028}, the pattern of P1, the bridge emission, P2 and
the off-pulse interval at phase 0.6 -- 1 can be well reproduced with
the model although the model underestimates the left wing of P1,
i.e., in the phase interval 0.0 -- 0.16.

\begin{figure}
\includegraphics[scale=0.8]{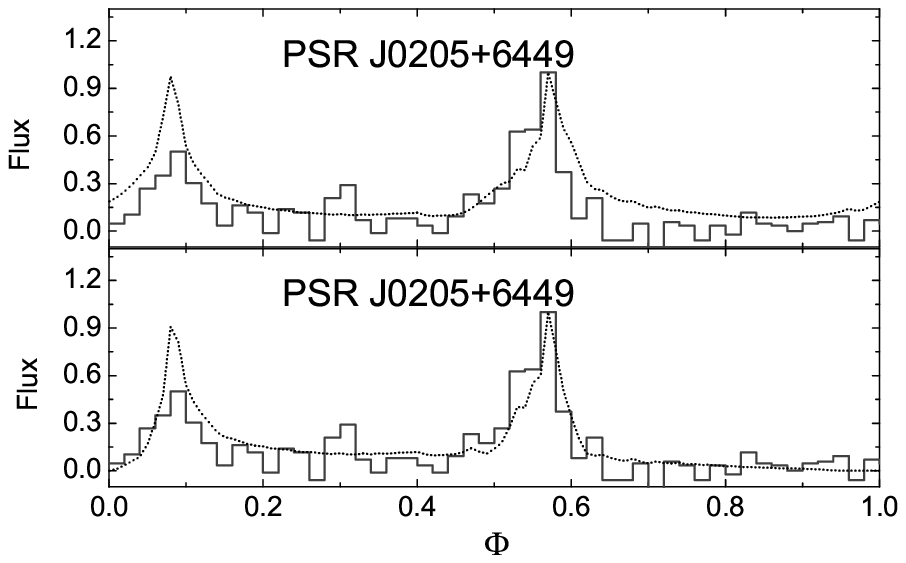}
\caption{\label{Fig0205} Comparison of the resulting light curve
with the {\it Fermi} observation ($>0.1$ GeV) \citep[][]{A09d} for
the pulsar PSR J0205+6449. Others are the same as
Fig.\ref{Figvela}.}
\end{figure}

The pulsar PSR J0205+6449 has a period of 65.7 ms and locates in the
PWN 3C58 with a distance of 3.2 kpc. Pulsed signals are first
discovered in X-rays from the {\it Chandra X-ray observatory} data
\citep[][]{wp78,bea82,rea93,mea02}. The X-ray profile shows two
narrow peaks separated by 0.5 in phase, and the first X-ray peak
lags the radio peak at 2.0 GHz  by $\sim 0.10$ in phase
\citep[][]{mea02,lea09}. This pulsar also was recently observed with
the LAT on {\it Fermi}, and 2922 effectual pulsed photons with
energies $>0.1$ GeV were detected \citep[][]{A09d}. The light curve
also shows two peaks separated by $0.49\pm0.02$ in phase. The first
peak P1 lags the radio pulse by $\sim0.08$ in phase, and the seconde
peak P2 is asymmetric with a slow rise and a fast fall
\citep[][]{A09d}.

The comparison of the {\it Fermi} LAT observation and the resulting
light curve for PSR J0205+6449 is shown in Fig.\ref{Fig0205}, and
the parameters are $P=65.7$ ms, $\alpha=85^\circ$, $\zeta_{\rm
obs}=75^\circ$, $\sigma=0.04$ and $\delta_{\rm ph} = 0.01 $. For
these parameters, the flux both at P1 and in the off-pulse region
(phase = 0.65 -- 1.0) is significantly reduced if emission from the
field lines in the emitting region with $r_{\rm nul}>0.9R_{\rm lc}$
is excluded, which is more consistent with the observation. With the
prescription taken into account, the model can broadly reproduce the
GeV light curve obtained with the LAT for pulsar PSR J1028-5819,
although it underestimates the flux in the left wing (phase = 0.0 --
0.08)(see Fig.\ref{Fig0205}).

\section{Discussion and conclusions}
\label{sec:discussion}

In this paper, we investigate the high-energy $\gamma$-ray light
curves from pulsars using a revised two-pole caustic model and apply
it to the four pulsars, Vela, PSR J1028-5819, PSR J0205+6449, and
PSR J2021+3651 whose light curves in the GeV band have been obtained
with the LAT on the {\it Fermi} telescope. The calculation basically
follows the two-pole caustic model proposed by \citet[][]{dr03} and
\citet[][]{Dea04}. The original two-pole caustic model assumes
high-energy particles are rim-dominated ($r_{\rm ovc}^0 = 1$) and
emit $\gamma$-rays along all the last open filed lines extending
from the stellar surface to high altitude, i.e., the polar cap
azimuth extension of the emitting region is $360^\circ$. However, in
our revised two-pole caustic model, the gap accelerating particles
is screened by the pair creation process along the azimuthal
direction, and the last open field lines are usually treated as a
boundary of the gap. As a result, we use $r_{\rm ovc}^0 = 0.98$ in
this paper.  Moreover, we find that the resulting high-energy light
curve for Vela is significantly higher than the {\it Fermi} result,
and this discrepancy can be eliminate if excluding the emission from
the field lines where $r_{\rm nul}>0.9R_{\rm lc}$. Physically, in
the former outer-gap model \citep[e.g.,][]{zc97,zc98,cet00}, the
inner boundary of the outer gap is the null charge surface where
pairs creation process is significant; more recent electrodynamical
studies show that if the current through the gap is taken into
account, the inner boundary can shift to the star surface, and the
location of the inner boundary is determined by the current injected
from the inner and outer boundaries and that produced in the gap
\citep[][]{hs01a,tsh04,h06,h08,tea08}. For the field lines with
relatively high null charge surface, the gap cannot be formed or the
gap is not effective in producing high-energy photons. However, the
details of the three-dimensional gap are usually uncertain for a
pulsar due to both the sensitivity of the pair creation process in
the gap and the unknown of the outer current. On average, it can be
assumed the gap cannot be formed on the field lines where $r_{\rm
nul}>0.9R_{\rm lc}$, and high-energy photons can be produced
effectively and uniformly from the star surface to high altitude
along the other field lines. Note that the electric field strength
near the inner boundary in these models is usually not big enough to
make the local particles radiation reaction-limited
\citep[][]{h06,h08,tea08}, and then the uniformity of the emissivity
of the high-energy photons from the star surface to high altitude is
violated. Therefore, our geometrical model for the light curves of
pulsars has a discrepancy with the outer-gap model.

Recently, \citet[][]{BS09a} argued that the treatment of the
aberration effect with the retarded dipole magnetic field in the
original TPC model is not self-consistent because the retarded
dipole magnetic field is valid in the lab frame rather than the
instantaneously corotating frame. We note that the retarded dipole
formula is an approximation to the more realistic force-free field
to the first order of $r/R_{LC}$, while the difference between
treating the retarded dipole field in different frames is to the
second order of $r/R_{LC}$. Moreover, the calculated high-energy
light curves are usually consistent with the observations when the
retarded dipole magnetic field is assumed to be valid in the
instantaneously corotating frame \citep[see also in][]{tea08,Vea09}.
Therefore, in our paper, a retarded vacuum dipole geometry of the
magnetic field around the pulsar is employed as in the
instantaneously corotating frame.

Profiles of $\gamma$-ray light curves of pulsars are sensitive to
the structure of the magnetic field lines and the location of the
gap, and uncertainties still exists using the vacuum magnetic field
since the magnetic filed structure near the light cylinder can be
significantly influenced by plasma current \citep[][]{BS09a}.
\citet[][]{BS09b} modeled $\gamma$-ray pulsar light curves with a
simulated force-free magnetic field. They found that double-peak
pulse profiles cannot be produced either in the conventional TPC
model or in the outer-gap scenario using the force-free structure.
Instead, an "annular gap" is proposed, in which the emission zone is
located at open field line regions that are just outside the current
sheet. Nevertheless, the origin of the gap and particle acceleration
mechanism are still lacking.

Photons with energies above several GeV can been attenuated either
through $\gamma$-$\gamma$ or through $\gamma$-B interaction in the
magnetosphere, and the profile of the light curve changes with
different energies. As a result, the model, which assumes the
high-energy photons are emitted uniformly along the field lines in
the gap, cannot be used to interpret the difference of the light
curves with different energies. On the other hand, the emissivity of
photons with low energies about several tens of MeV changes greatly
in the magnetosphere, and then the model also cannot be employed to
study the MeV emission of pulsar. However, because the electrons can
be radiation reaction-limited within a large range of altitudes in
the gap \cite[][]{hhs03,dr03}, the assumption is reasonable for the
GeV photons which do not encountered significant attenuation.

The four pulsars, Vela, PSR J1028-5819, PSR J0205+6449, and PSR
J2021+3651, have been observed with the LAT on {\it Fermi}. The GeV
light curves for the four pulsars are usually shown with two main
peaks with phase intervals from 0.43 -- 0.5. With appropriate
parameters, the GeV light curves can be broadly reproduced with the
model. For Vela, PSR J1028-5819 and PSR J2021+3651, the emission
level in the off-pulse region is greatly reduced excluding the
emission from the field lines with $r_{\rm nul}>0.9R_{\rm lc}$,
which is more consistent with the observations.

\section*{Acknowledgments}
This work is partially supported by the National Natural Science
Foundation of China (NSFC 10778702, 10803005), a 973 Program
(2009CB824800), and Yunnan Province under a grant 2009 OC.

%%%%%%%%%%%%%%%%%%%%%%%%%%%%%%%%%%%%%%%%%%%%%%%%%%%%%%%%%%%%%%%%%%%%%%%%%%%%%%%
%%%%%%%%%%%%%%%%%%%%%%%%%%%%% BIBLIOGRAPHY %%%%%%%%%%%%%%%%%%%%%%%%%%%%%%%%%%%%
%%%%%%%%%%%%%%%%%%%%%%%%%%%%%%%%%%%%%%%%%%%%%%%%%%%%%%%%%%%%%%%%%%%%%%%%%%%%%%%

\end{document}